\documentclass[12pt,journal,compsoc]{IEEEtran}

\ifCLASSOPTIONcompsoc
\else
\fi
\usepackage{multirow}

\ifCLASSINFOpdf
  \usepackage[pdftex]{graphicx}
\else
\fi
\usepackage{hyperref}

\pdfoutput=1
\begin{document}
%
\title{A Study of Medium Access Control Protocols for Wireless Body Area Networks}
%
%
%
%

\author{Sana~Ullah$^{\Psi}$, Bin~Shen, S.M.Riazul~Islam, Pervez Khan, Shahnaz Saleem, and~Kyung Sup~Kwak
\IEEEcompsocitemizethanks{\IEEEcompsocthanksitem S.Ullah, B.Shen, S.M.R. Islam, P. Khan, and K.S. Kwak are with the Graduate School of Telecommunication Engineering, Inha University, Incheon (402-751)
South Korea.\protect\\
\IEEEcompsocthanksitem $\Psi$. Author to whom correspondence should be addressed E-mail: sanajcs@hotmail.com \protect\\
\IEEEcompsocthanksitem S. Saleem is with the Graduate School of Computer Science and Engineering, Inha University, Incheon (402-751) South Korea.}
\thanks{Manuscript received November 26, 2009; revised December 7, 2009; published (Sensors, vol. 10, no. 1: 128-145.) December 20, 2009.}}

\IEEEcompsoctitleabstractindextext{%
\begin{abstract}
The seamless integration of low-power, miniaturised, invasive/non-invasive lightweight sensor nodes have contributed to the development of a proactive and unobtrusive Wireless Body Area Network (WBAN). A WBAN provides long-term health monitoring of a patient without any constraint on his/her normal dailylife activities. This monitoring requires low-power operation of invasive/non-invasive sensor nodes. In other words, a power-efficient Medium Access Control (MAC) protocol is required to satisfy the stringent WBAN requirements including low-power consumption. In this paper, we first outline the WBAN requirements that are important for the design of a low-power MAC protocol. Then we study low-power MAC protocols proposed/investigated for WBAN with emphasis on their strengths and weaknesses. We also review different power-efficient mechanisms for WBAN. In addition, useful suggestions are given to help the MAC designers to develop a low-power MAC protocol that will satisfy the 
stringent WBAN requirements.
\end{abstract}

}

\maketitle

\IEEEdisplaynotcompsoctitleabstractindextext

%
\IEEEpeerreviewmaketitle

\section{introduction}
A Wireless Body Area Network (WBAN) allows the integration of intelligent, miniaturized, low- power, invasive/non-invasive sensor nodes in/around a  human body that are used to monitor body functions and the surrounding environment. Each intelligent node has enough capability to process and forward information to a base station for diagnosis and prescription. A WBAN provides long term health monitoring of patients under natural physiological states without constraining their normal activities. It is used to develop a smart and affordable health care system and can be a part of diagnostic procedure, maintenance of a chronic condition, supervised recovery from a surgical procedure, and can handle emergency events. 

Some of the common objectives in WBAN are to achieve maximum throughput, minimum delay, and to maximize the network lifetime by controlling the main sources of energy waste, i.e., collision, idle listening, overhearing, and control packet overhead. A collision occurs when more than one packet is transmitted at the same time. The collided packets have to be retransmitted, which consumes extra energy. The second source of energy waste is idle listening, meaning that a node listens to an idle channel to receive data. The third source is overhearing, i.e., to receive packets that are destined to other nodes. The last source is control packet overhead, meaning that control information are added to the payload. A minimal number of control packets should be used for data transmission. Medium Access Control (MAC) protocols play an important role in solving the aforementioned problems. Generally they are grouped into contention-based and schedule-based MAC protocols. In contention-based MAC protocols such as Carrier Sense Multiple Access/Collision Avoidance (CSMA/CA) protocols, nodes contend for the channel to transmit data. If the channel is busy, the node defers its transmission until it becomes idle. These protocols are scalable with no strict time synchronization constraint. However, they incur significant protocol overhead. In schedule-based protocols such as Time Division Multiple Access (TDMA) protocols, the channel is divided into time slots of fixed or variable duration. These slots are assigned to nodes and each node transmits during its slot period. These protocols are energy conserving protocols. The duty cycle of radio is reduced and there is no contention, idle listening and overhearing problems. These protocols, however, require frequent synchronization. Table \ref{tab:1} compares CSMA/CA and TDMA protocols.

\begin{table}[!t]
\renewcommand{\arraystretch}{1.3}
\caption{CSMA vs. TDMA Protocols}
\label{tab:1}
\centering
\begin{tabular}{|c|c|c|}
\hline
Performance Metric &	CSMA/CA &	TDMA\\
\hline
Power consumption &	High &	Low \\
\hline
Traffic level &	Low &	High\\
\hline
Bandwidth utilisation &	Low	& Maximum\\
\hline
Scalability &	Good &	Poor\\
\hline
Effect of packet failure &	Low &	Latency\\
\hline
Synchronisation	& Not Applicable &	Required\\
\hline
\end{tabular}
\end{table}

The development of a low-power MAC protocol for WBAN has been a hot research topic for the last few years. Considerable research efforts have been dedicated to propose/investigate new MAC protocols that could satisfy the crucial WBAN requirements. A number of researchers have considered IEEE 802.15.4 \cite{1} for WBAN since it supports low data rate applications, but it is not enough to support high data rate applications (data rate $>$ 250 Kbps). Other protocols such as Heartbeat Driven MAC (H-MAC) \cite{2}, Reservation-based Dynamic TDMA (DTDMA) \cite{3}, Preamble-based TDMA (PB-TDMA) \cite{4}, and BodyMAC \cite{5} protocols have been proposed/investigated in the existing literature. In this paper, an overview of the aforesaid protocols with focus on their strengths and weaknesses is presented. Useful suggestions are given to overcome the weaknesses of these protocols. Then, a comparative analysis of many power-efficient mechanisms such as Low-power Listening (LPL), scheduled-contention, and TDMA mechanisms is presented in the context of WBAN. Examples are given to validate the discussion.

The rest of the paper is categorized into four sections. Section 2 presents the WBAN requirements. A study of different low-power MAC protocols proposed/investigated for WBAN is given in Section 3. Section 4 reviews various power-efficient mechanisms for WBAN with useful guidelines. The final section concludes our work. 

\section{wban mac requirements}
The most important attribute of a good MAC protocol for WBAN is energy efficiency. In some applications, the device should support a battery life of months or years without intervention, while others may require a battery life of only tens of hours due to the nature of the applications. For example, cardiac defibrillators and pacemakers should have a lifetime of more than 5 years, while swallowable camera pills have a lifetime of 12 hours \cite{6}. Power-efficient and flexible duty cycling techniques are required to minimize the idle listening, overhearing, packet collisions and control packet overhead problems. Furthermore, low duty cycle nodes should not receive frequent synchronization and control information (beacon frames) if they have no data to send or receive. 

\begin{figure*}[!t]
\centering
\includegraphics[width=5in]{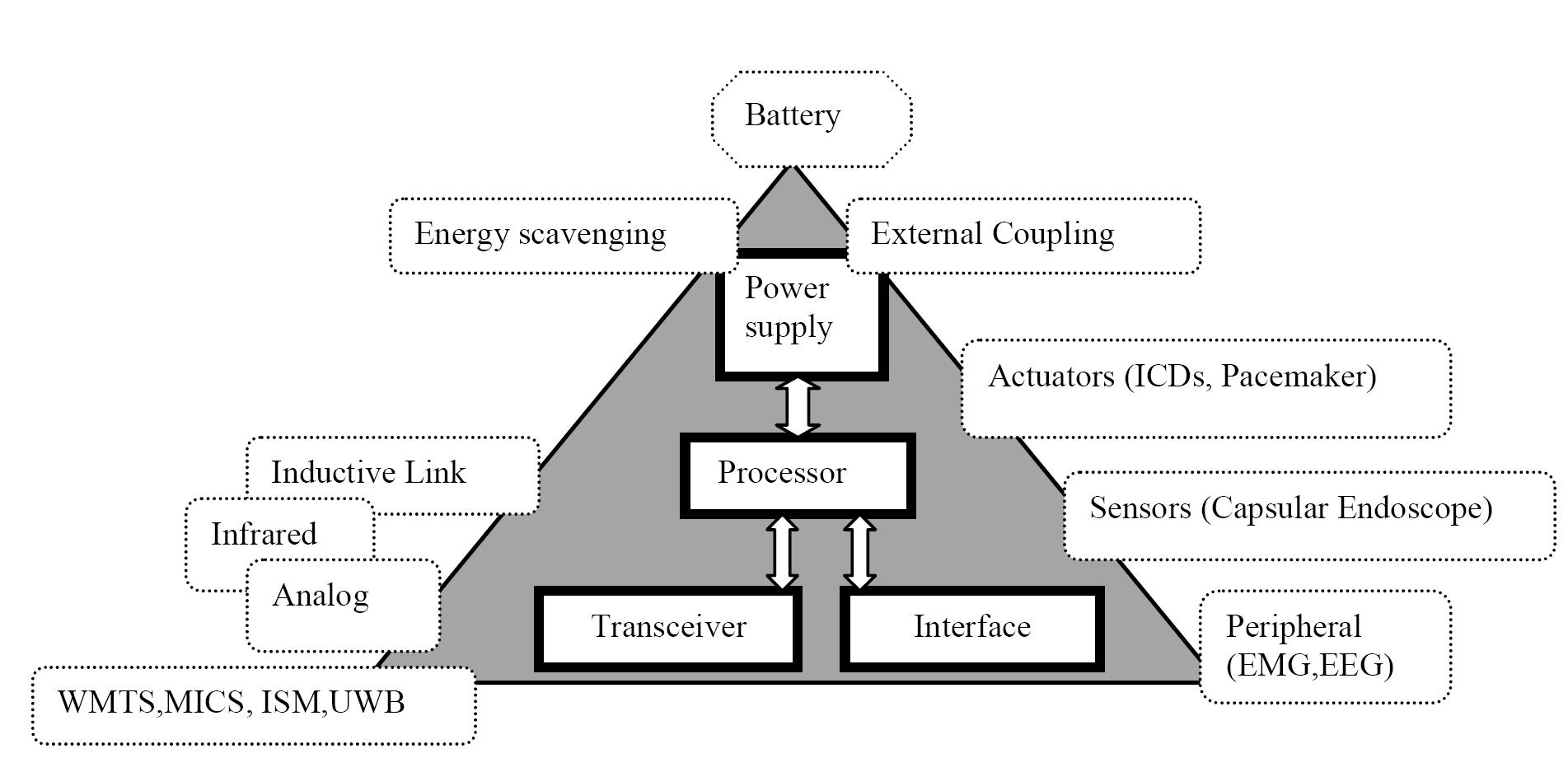}
\caption{Potential issues of a MAC protocol for WBAN}
\label{fig:1}
\end{figure*}

The WBAN MAC should support simultaneous operation on in-body (Medical Implant Communications Service, also called MICS) and on-body frequency bands/channels [Industrial, Scientific and Medical (ISM) or Ultra Wide Band (UWB)] at the same time. In other words, it should support Multiple Physical layers (Multi-PHYs) communication. Other important factors are scalability and adaptability to changes in the network, delay, throughput, and bandwidth utilization. Changes in the network topology, the position of the human body, and the node density should be handled rapidly and successfully. The MAC protocol for WBAN should consider the electrical properties of the human body and the diverse traffic nature of in-body and on-body nodes. For example, the data rate of in-body nodes varies, ranging from few kbps in pacemaker to several Mbps in capsular endoscope. Fig. \ref{fig:1} shows some of the potential issues of a MAC protocol for WBANs. 

The Quality of Service (QoS) is also an important factor of a good MAC protocol for WBAN. This includes point-to-point delay and delay variation. In some cases, real-time communication is required for many applications such as fitness and medical surgery monitoring applications. For multimedia applications, the latency should be less than 250 ms and the jitter should be less than 50 ms \cite{6}. However, the reliability, latency, and jitter requirements depend on the nature of the applications. For emergency applications, the MAC protocol should allow in-body or on-body nodes to get quick access to the channel (in less than one second) and to send the emergency data to the coordinator. One such example is the detection of irregular heartbeat, high or low blood pressure or temperature, and excessively low or high blood glucose level in a diabetic patient. Another example is when the node is dying. Reporting medical emergency events should have higher priority than non-medical emergency (battery dying) events.   

In WBAN, most of the traffic is correlated, i.e., a patient suffering from fever triggers temperature, blood pressure, and respiration sensors at the same time \cite{7}. These changes may also affect the oxygen saturation level (SpO2) in the blood. These kinds of physiological parameters increase the traffic correlation. A single physiological fluctuation triggers many sensors at the same time. In this case, a CSMA/CA protocol encounters heavy collisions and extra energy consumption. Also, in CSMA/CA protocol, the nodes are required to perform Clear Channel Assessment (CCA) before transmission. However, the CCA is not always guaranteed in the MICS band since the path loss inside the human body due to tissue heating is much higher than in free space. Bin et.al studied the unreliability of a CCA in WBAN and concluded that for a given -85 dBm CCA threshold, the on-body nodes cannot see the activity of the in-body nodes when they are at 3 meters distance away from the surface of the body \cite{8}. Sana et al. have studied the behavior of the CSMA/CA protocol for WBAN and concluded that the CSMA/CA protocol encounters heavy collision problems for high traffic nodes \cite{9}. TDMA-based protocols provide good solutions to the traffic correlation, heavy collision, and CCA problems. These protocols are energy conserving protocols because the duty cycle is reduced and there are no contention, idle listening, and overhearing problems. However, common TDMA needs extra energy for periodic time synchronization. All the sensors (with and without data) are required to receive the periodic packets in order to synchronize their clocks. Therefore, the design and implementation of a new TDMA protocol is required that can accommodate the heterogeneous WBAN traffic in a power-efficient manner.
\begin{table*}[!t]
\renewcommand{\arraystretch}{1.3}
\caption{IEEE 802.15.4 frequency bands, data rates, and modulation methods}
\label{tab:2}
\centering
\begin{tabular}{|c|c|c|c|c|c|}
\hline
\bf Frequency Bands/Channels & \bf Coverage & \bf Sub-channels & \bf Data Rate & \bf Data Modulation & \bf Chip Modulation\\
\hline
 2.4 GHz & Worldwide & 16 & 250kbit/s & 16-ary orthogonal & OQPSK,2Mchips/s\\
\hline
868 MHz & Europe & 1 & 20kbit/s & BPSK & BPSK,300kchips/s\\
\hline
915 MHz & Americas & 10 & 40kbit/s & BPSK & BPSK,600kchips/s\\

\hline
\end{tabular}
\end{table*}
\section{existing/proposed mac protocols for wban}

\subsection{IEEE 802.15.4}

\begin{figure}[!h]
\centering
\includegraphics[width=3.6in]{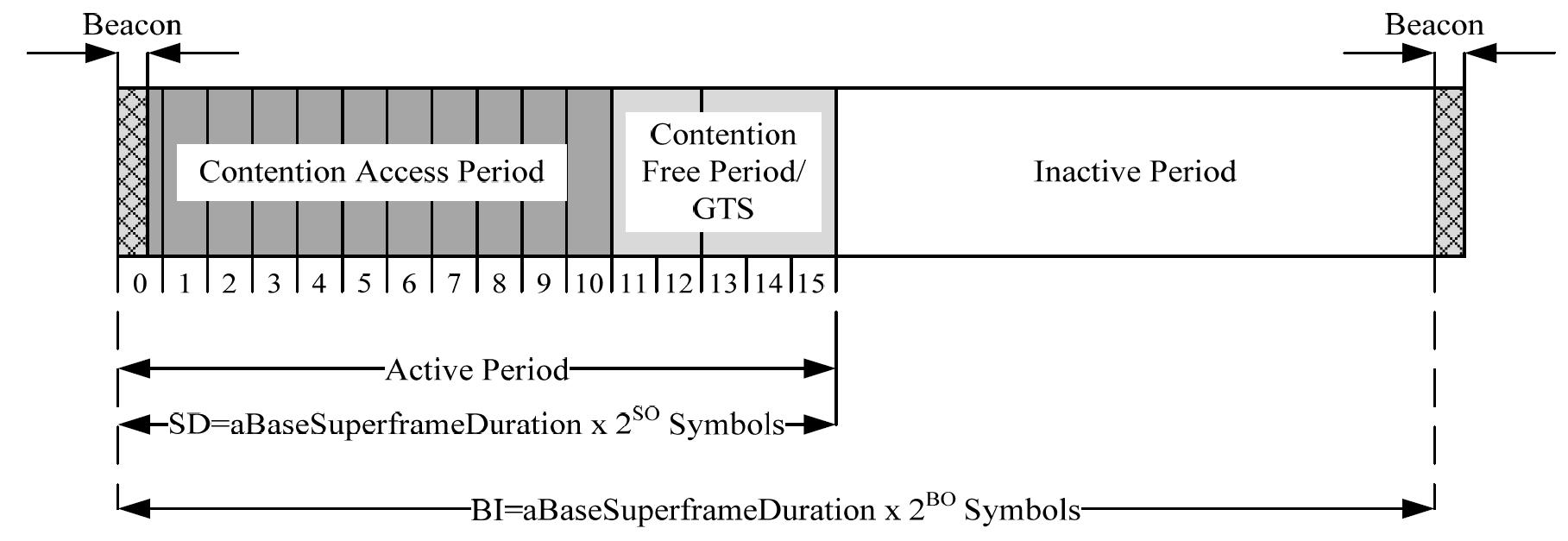}
\caption{IEEE 802.15.4 superframe structure}
\label{fig:2}
\end{figure}

IEEE 802.15.4 is a low-power standard designed for low data rate applications. It offers three operational frequency bands: 868 MHz, 915 MHz, and 2.4 GHz bands. There are 27 sub-channels allocated in IEEE 802.15.4, i.e., 16 sub-channels in 2.4 GHz band, 10 sub-channels in 915 MHz band and one sub-channel in the 868 MHz band, as given in Table \ref{tab:2}. The table also shows the data rate and the modulation technique for each frequency band. IEEE 802.15.4 has two operational modes: a beacon-enabled mode and a non-beacon enabled mode. In a beacon-enabled mode, the network is controlled by a coordinator, which regularly transmits beacons for device synchronization and association control. The channel is bounded by a superframe(s) structure as illustrated in Fig. \ref{fig:2}.  

The superframe consists of both active and inactive periods. The active period contains three components: a beacon, a Contention Access Period (CAP), and a Contention Free Period (CFP). The length of the entire superframe (Beacon Interval, BI) and the length of the active part of the superframe (Superframe Duration, SD) are defined as follows

\begin{equation}
BI=aBaseSuperframeDuration¡¿2^{BO}
\end{equation}
\begin{equation}
SD=aBaseSuperframeDuration¡¿2^{SO} 
\end{equation}

where aBaseSuperframeDuration=960 symbols, BO=Beacon Order, and SO=Superframe Order. The coordinator interacts with nodes during the active period and sleeps during the inactive period. There are maximum seven Guaranteed Time Slots (GTSs) in the CFP period to support time critical traffic. In a beacon-enabled mode, a slotted CSMA/CA protocol is used in the CAP period while in a non-beacon enabled mode, an unslotted CSMA/CA protocol is used.
\begin{table*}[!t]
\renewcommand{\arraystretch}{1.3}
\caption{Throughput at a 0 dBm Transmit Power \cite{12}}
\label{tab:3}
\centering
\begin{tabular}{c|c|c|c|c|c|c|l}
\cline{2-7}
& \multicolumn{3}{|c|}{Person is Standing (Destination Nodes)} & \multicolumn{3}{|c|}{Person is Sitting (Destination Nodes)} \\ \cline{2-7}
& Chest & Waist & Ankle & Chest & Waist & Ankle \\ \cline{1-7}
\multicolumn{1}{|c|}{Chest} & - & 99\% & 84\% & - & 99\% & 81\%    \\ \cline{1-7}
\multicolumn{1}{|c|}{Waist} & 100\% & - & 50\% & 99\% & - & 47\%    \\ \cline{1-7}
\multicolumn{1}{|c|}{Ankle} & 72\% & 76\% & - & 77\% & 27\% & -    \\ \cline{1-7}
\end{tabular}
\end{table*}

\begin{table*}[!t]
\renewcommand{\arraystretch}{1.3}
\caption{Throughput and Power (in mW) of IEEE 802.15.4 and IEEE 802.11e under AC\textunderscore BE and AC\textunderscore VO \cite{15}}
\label{tab:4}
\centering
\begin{tabular}{|c|c|c|c|c|}
\cline{1-5}
& \multicolumn{1}{|c|}{} & \multicolumn{1}{|c|}{IEEE 802.15.4)} & \multicolumn{1}{|c|}{IEEE 802.11e (AC\textunderscore BE)
)} & \multicolumn{1}{|c|}{IEEE 802.11e (AC\textunderscore VO))} \\ \cline{1-5}

\multicolumn{1}{|c|}{Throughput} & Wave-form  & 100\% & 100\% & 100\%       \\ \cline{2-5}
\multicolumn{1}{|c|}{} & Parameter & 99.77\% & 100\% & 100\%       \\ \cline{1-5}
\multicolumn{1}{|c|}{Power (mW)} &  Wave-form & 1.82 & 4.01 & 3.57       \\ \cline{2-5}
\multicolumn{1}{|c|}{} &  Parameter & 0.26 & 2.88 & 2.77       \\ \cline{1-5}
\end{tabular}
\end{table*}

IEEE 802.15.4 has remained the main focus of many researchers during the past few years. Some of the main reasons of selecting IEEE 802.15.4 for WBAN are low-power communication and support of low data rate WBAN applications. Nicolas et al. investigated the performance of a non-beacon IEEE 802.15.4 in \cite{10}, where low upload/download rates (mostly per hour) are considered. They concluded that the non-beacon IEEE 802.15.4 results in 10 to 15 years sensor lifetime for low data rate and asymmetric WBAN traffic. However, their work considered data transmission on the basis of periodic intervals, which is not a perfect scenario in a real WBAN. Furthermore, the data rate of in-body and on-body nodes is not always low, i.e., it ranges from 10 Kbps to 10 Mbps, and hence reduces the lifetime of the sensor nodes. Li et al. studied the behaviour of slotted and unslotted CSMA/CA protocols and concluded that the unslotted mechanism performs better than the slotted one in terms of throughput and latency, but with a high power consumption cost \cite{11}.

Intel Corporation conducted a series of experiments to analyze the performance of IEEE 802.15.4 for WBAN \cite{12}. They deployed a number of Intel Mote 2 \cite{13} nodes on chest, waist, and the right ankle. Table \ref{tab:3} shows the throughput at a 0 dBm transmit power when a person is standing and sitting on a chair. The connection between ankle and waist cannot be established, even for a short distance of 1.5 m. All other connections show favourable performance. 

Dave et al. studied the energy efficiency and QoS performance of IEEE 802.15.4 and IEEE 802.11e \cite{14} MAC protocols under two generic applications: a wave-form real time stream and a real-time parameter measurement stream \cite{15}. Table \ref{tab:4} shows the throughput and the Power (in mW) for both applications. The AC\textunderscore BE and AC\textunderscore VO represent the access categories best-effort and voice in the IEEE 802.11e, respectively.

Since the IEEE 802.15.4 operates in the 2.4 GHz unlicensed band, the possibilities of interference from other devices such as IEEE 802.11 and microwave are inevitable. A series of experiments to evaluate the impact of IEEE 802.11 and microwave ovens on the IEEE 802.15.4 transmission are carried out in \cite{16}. The authors considered XBee 802.15.4 development kit having two XBee modules. Table \ref{tab:5} shows the effects of microwave oven on the XBee remote module. When the microwave oven is ON, the packet success rate and the standard deviation (Std.) is degraded to 96.85\% and 3.22\% respectively. However, there is no loss when the XBee modules are taken two meters away from the microwave oven.

\begin{table*}[!t]
\renewcommand{\arraystretch}{1.3}
\caption{Co-existence Test Results between IEEE 802.15.4 and Microwave Oven \cite{16}}
\label{tab:5}
\centering
\begin{tabular}{|c|c|c|}
\hline
Microwave Status & Packet Success Rate (Mean) & Packet Success Rate (Std.) \\
\hline
ON & 96.85\% & 3.22\% \\
\hline
OFF & 100\% & 0\% \\
\hline
\end{tabular}
\end{table*}

\subsection{Heartbeat Driven MAC (H-MAC) Protocol}
A Heartbeat Driven MAC protocol (H-MAC) is a TDMA-based protocol originally proposed for a star topology WBAN. The energy efficiency is improved by exploiting heartbeat rhythm information in order to synchronize the nodes. The nodes do not need to receive periodic information to perform synchronization. The heartbeat rhythm can be extracted from the sensory data and hence all the rhythms represented by peak sequences are naturally synchronized. The H-MAC protocol assigns dedicated time slots to each node to guarantee collision-free transmission. This protocol is supported by an active synchronization recovery scheme supported by two resynchronization schemes. 

Although H-MAC protocol reduces extra energy cost required for synchronization, it does not support sporadic events. Since the TDMA slots are dedicated and are not traffic adaptive, H-MAC protocol encounters low spectral/bandwidth efficiency in case of low traffic. For example, a blood pressure node may not need a dedicated time slot while an endoscope pill may require a number of dedicated time slots when deployed in WBAN. But the slots should be released when the endoscope pill is expelled. The heartbeat rhythm information varies depending on the patient condition. It may not reveal valid information for synchronization all the time. One of the solutions is to assign the time slots based on the node's traffic information and to receive synchronization packets when required, i.e., when a node has data to transmit/receive.

\subsection{Reservation-based Dynamic TDMA (DTDMA) Protocol}
\begin{figure}[!h]
\centering
\includegraphics[width=3.5in]{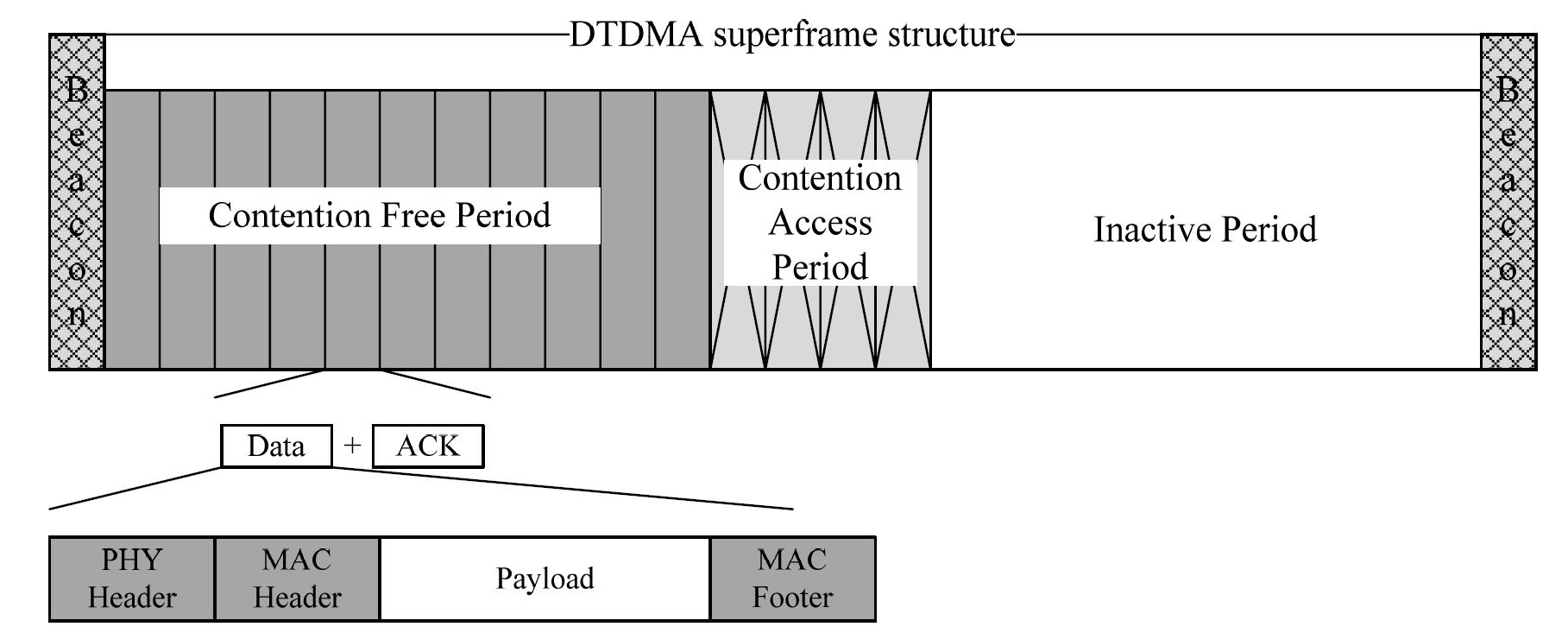}
\caption{DTDMA superframe structure}
\label{fig:3}
\end{figure}

A Reservation-based Dynamic TDMA Protocol (DTDMA) is originally proposed for normal (periodic) WBAN traffic where slots are allocated to the nodes having buffered packets and are released to other nodes when the data transmission/reception is completed. The channel is bounded by superframe structures. Each superframe consists of a beacon - used to carry control information including slot allocation information, a CFP period - a configurable period used for data transmission, a CAP period - a fixed period used for short command packets using slotted-ALOHA protocol, and a configurable inactive period - used to save energy. Unlike a beacon-enabled IEEE 802.15.4 superframe structure where the CAP duration is followed by CFP duration, in DTDMA protocol the CFP duration is followed by CAP duration in order to enable the nodes to send CFP traffic earlier than CAP traffic. In addition, the duration of an inactive period is configurable based on the CFP slot duration. If there is no CFP traffic, the inactive period will be increased. The DTDMA superframe structure is given in Fig. \ref{fig:3}.

It has been shown that for normal (periodic) traffic, the DTDMA protocol provides more dependability in terms of low packet dropping rate and low energy consumption when compared with IEEE 802.15.4. However, it does not support emergency and on-demand traffic. Although the slot allocation based on the traffic information is a good approach, the DTDMA protocol has several limitations when considered for the MICS band. The MICS band has ten sub-channels where each sub-channel has 300 Kbps bandwidth. The DTDMA protocol could operate on one sub-channel but cannot operate on ten sub-channels simultaneously. In addition, the DTDMA protocol does not support the sub-channel allocation mechanism in the MICS band. This protocol can be further investigated for the MICS band by integrating the sub-channel information in the beacon frame. The new concept may be called Frequency-based DTDMA (F-DTDMA), i.e., the coordinator first selects one of the sub-channels in the MICS band and then divides the selected sub-channel in TDMA superframe(s) according to the DTDMA protocol. However the Federal Communications Commission(FCC) has imposed several restrictions on the sub-channel selection/allocation mechanism in the MICS band (see Section 4.4), which further creates problems for the MAC designers.

\subsection{PB-TDMA Protocol}
The performance of a Preamble-based TDMA (PB-TDMA, see Section 4.3) for WBAN has been analyzed in \cite{17}. The authors used NS-2 for extensive simulations where the wireless physical parameters were considered according to low-power Nordic nRF2401 transceiver and the simulation area was $3 \times 3$ meters. For the performance comparison, many other protocols such as S-MAC and IEEE 802.15.4 were also simulated in the same environment. Simulation results showed that PB-TDMA protocol outperformed S-MAC and IEEE 802.15.4 protocol in terms of energy efficiency. The results are valid for normal traffic only. The co-existence of emergency and on-demand traffic is not considered in the simulation. 

\subsection{BodyMAC Protocol}
\begin{figure}[!h]
\centering
\includegraphics[width=3.5in]{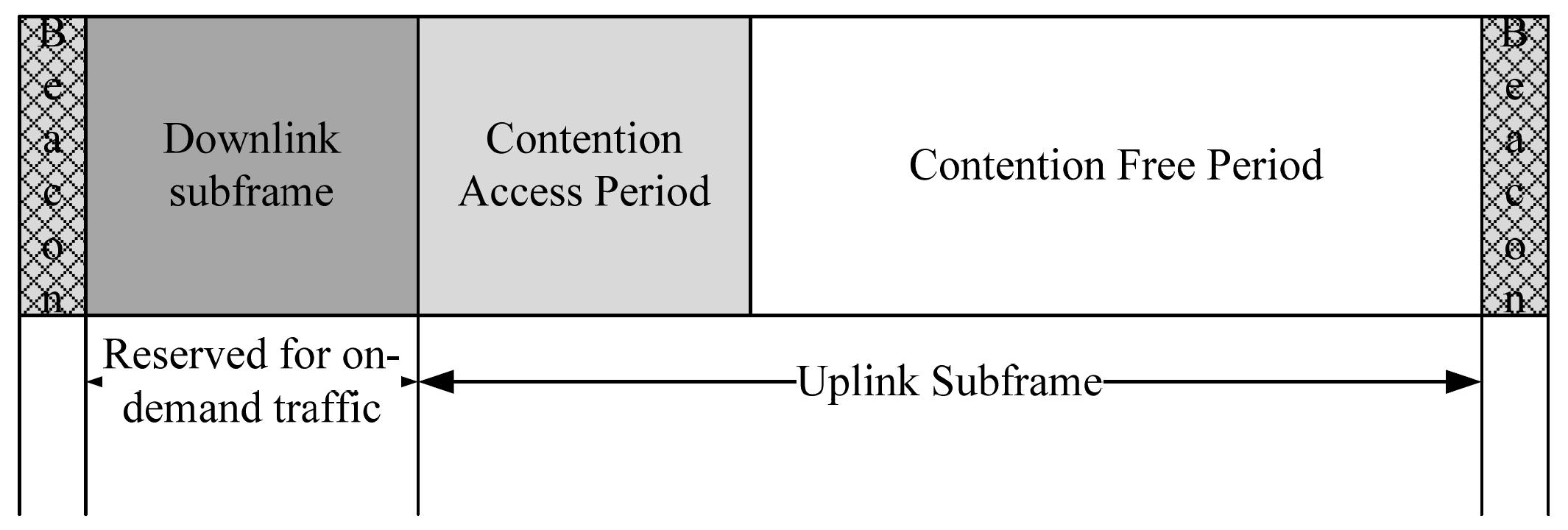}
\caption{BodyMAC superframe structure}
\label{fig:4}
\end{figure}

A BodyMAC protocol is a TDMA-based protocol where the channel is bounded by TDMA superframe structures with downlink and uplink subframes as given in Fig. \ref{fig:4}. The downlink frame is used to accommodate the on-demand traffic and the uplink frame is used to accommodate the normal traffic. There is no proper mechanism to handle emergency traffic. The uplink frame is further divided into CAP and CFP periods. The CAP period is used to transmit small size MAC packets. The CFP period is used to transmit the normal data in a TDMA slot. The duration of the downlink and uplink superframes are defined by the coordinator. 

The advantage of the BodyMAC protocol is that it accommodates the on-demand traffic using the downlink subframe. However, in case of low-power implants (which should not receive beacons periodically), the coordinator has to wake up the implant first and then send synchronization packets. After synchronization, the coordinator can request/send data in the downlink subframe. The wake up procedure for low-power implants is not defined in the BodyMAC protocol. One of the solutions is to use a wakeup radio in order to wake up low-power implants before using the downlink subframe. In addition the wakeup packets can be used to carry control information such as sub-channel (MICS band) and slot allocation information from the coordinator to the nodes. Finally, the BodyMAC protocol uses the CSMA/CA protocol in the CAP period which is not reliable for WBAN. This should be replaced by slotted-ALOHA as done in DTDMA.

\section{power-efficient mechanisms for wban}
Power-efficient mechanisms play an important role in the performance of a good MAC protocol. These mechanisms are categorized into Low-power Listening (LPL), Scheduled Contention, and TDMA mechanisms. The following sections briefly explain each mechanism with examples.

\subsection{Low-power Listening (LPL) Mechanism}
In a Low-power Listening (LPL) mechanism, nodes wake up for a short duration to check the channel activity without receiving any data. If the channel is idle the nodes go into sleep mode, otherwise they stay on the channel to receive the data. This is also called channel polling. The LPL is performed on regular basis regardless of synchronization among nodes. The sender sends a long preamble before each message in order to detect the polling at the receiving end.

\begin{figure}[!h]
\centering
\includegraphics[width=3.5in]{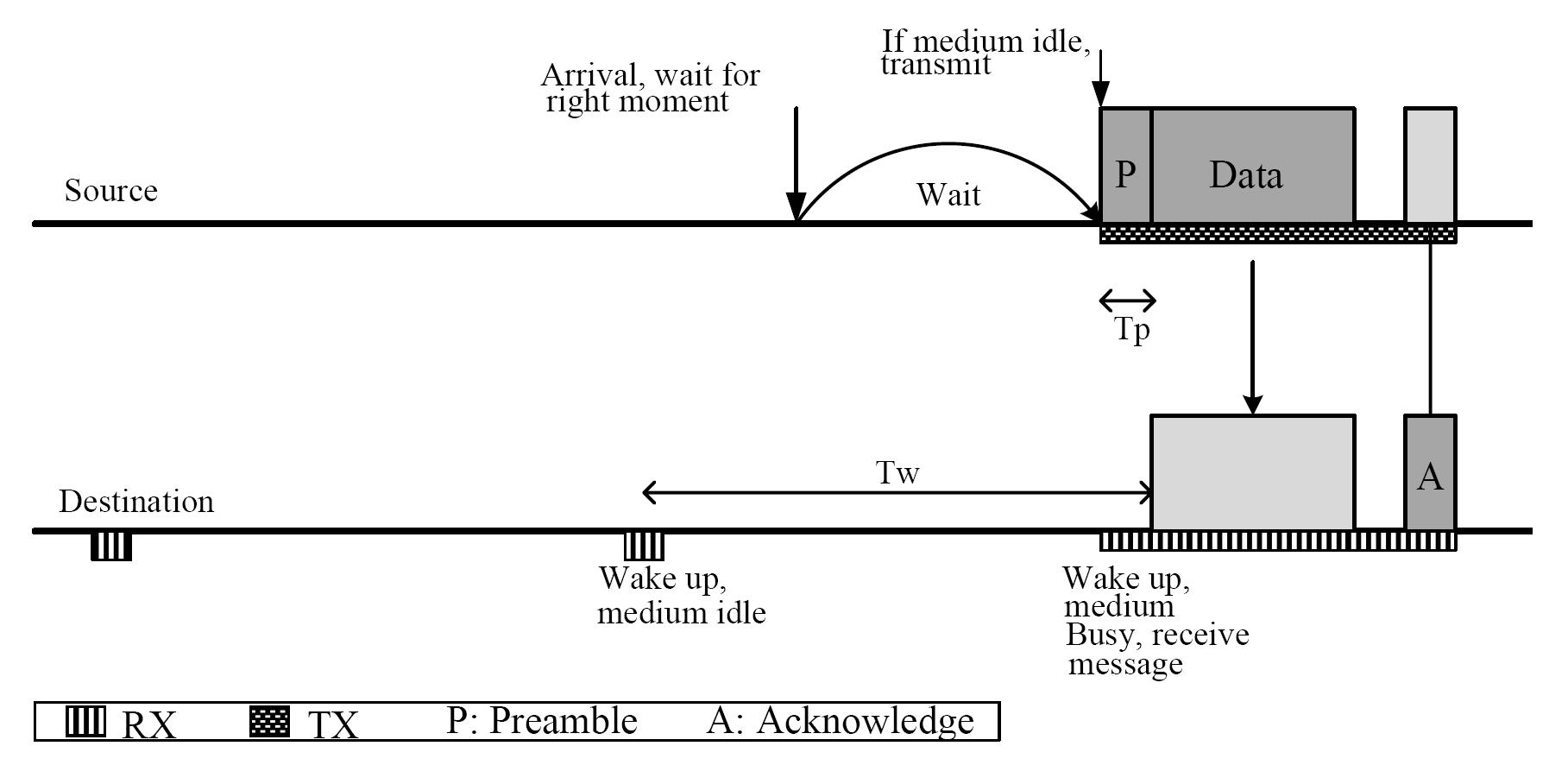}
\caption{WiseMAC concept}
\label{fig:5}
\end{figure}

The WiseMAC \cite{18} protocol is based on the LPL mechanism. In this protocol, a non-persistent CSMA and a preamble sampling technique is used to reduce idle listening. The preamble is used to alert the receiving node of a packet arrival. All the nodes in a network sample the medium periodically. If a node samples a busy medium, it continues to listen until it receives data or the medium becomes idle. Fig. \ref{fig:5} shows the WiseMAC concept.

In WBAN, the LPL mechanism has several advantages and disadvantages. The periodic sampling is efficient for high-traffic nodes and performs well under variable traffic conditions. However, it is ineffective for low-traffic nodes, especially in-body nodes, where periodic sampling is not preferred due to strict power constraints. Since the WBAN topology is a star topology and most of the traffic is uplink, using LPL mechanism is not an optimal solution to support both in-body and on-body communication simultaneously. 

\subsection{Scheduled-Contention Mechanism}
In a scheduled-contention mechanism, scheduled and contention based schemes are combined to incur scalability and collision avoidance. In this mechanism, the nodes adapt a common schedule for data communication. The schedules are exchanged periodically during a synchronization period. If two neighbouring nodes reside in two different clusters, they keep the schedules of both clusters resulting in extra energy consumption. 

\begin{figure}[!h]
\centering
\includegraphics[width=3.5in]{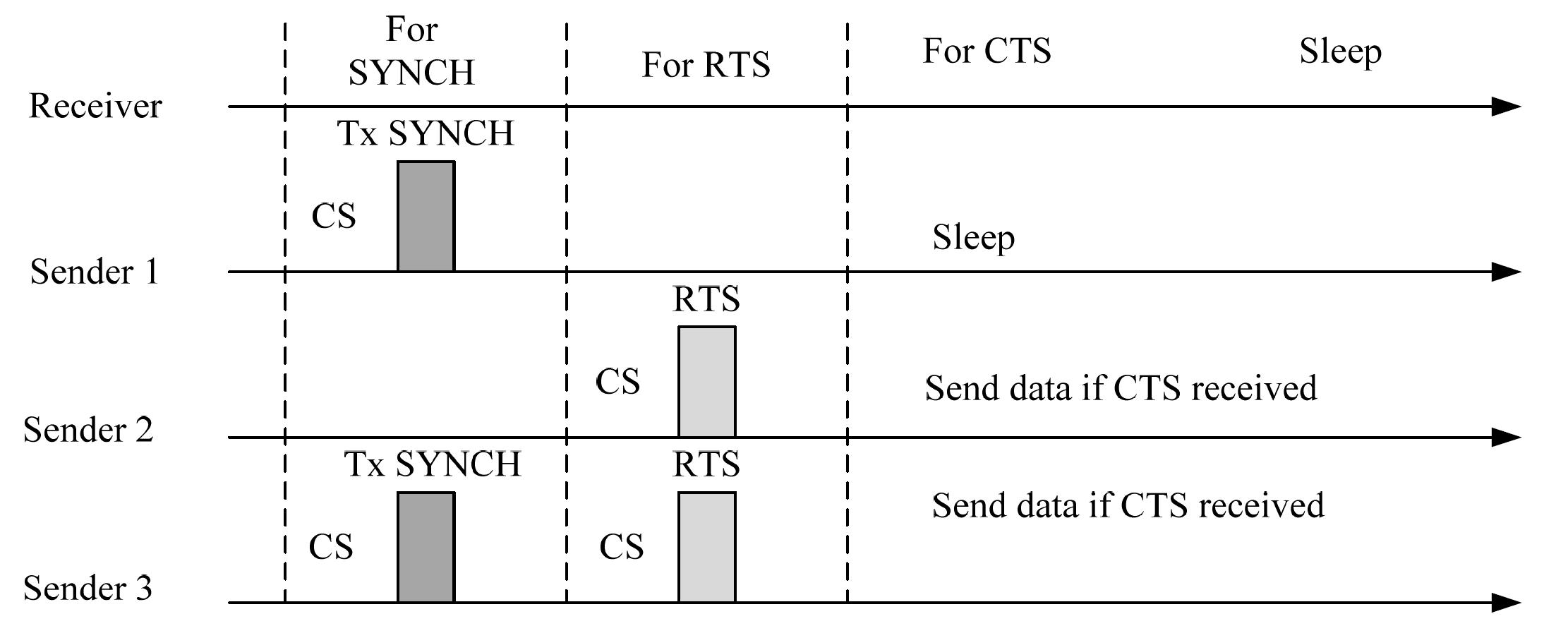}
\caption{Timing relationship between a receiver and different senders. CS is carrier sense.}
\label{fig:6}
\end{figure}

The S-MAC \cite{19} protocol is a good example of a scheduled-contention mechanism. S-MAC is a power-efficient and a contention-based MAC protocol designed for multi-hop Wireless Sensor Networks (WSNs). In this protocol, the low duty cycle mode is default operation of all nodes. This protocol introduces the concept of coordinated sleeping among neighbouring nodes. The node is active when it has data to send otherwise its radio is completely turned off. The energy is reduced from all the sources of energy waste, i.e., idle listening, collision, overhearing and control overhead. A complete cycle of listen and sleep is called frame. Each frame begins with a wakeup period, which is used by nodes to exchange control information. The wakeup period is usually followed by a sleep period. If a node has data to send while in the sleep mode, it must defer its transmission until the next wakeup slot. The state of the channel is determined using physical and virtual carrier sense mechanism. For each unicast frame transmission, Request to Send/Clear to Send (RTS/CTS) mechanism is followed. Broadcast frames are sent without using RTS/CTS mechanism. If a node fails to access the medium, it turns off its radio until the Network Allocation Vector (NAV) is zero. Nodes maintain synchronization by sending SYNCH packet. The size of SYNCH packet is very short and includes information about next sleep period. The listen period of a node is divided into two parts when both SYNCH and data packets are received at the same time. Fig. \ref{fig:6} illustrates the timing relationship between a receiver and different senders. Sender 1 sends a SYNCH packet only. Sender 2 sends a unicast data packet only. Sender 3 sends both a SYNCH and a data packet. 

A scheduled-contention mechanism reduces idle listening using sleep schedules and performs well for multi-hop WSNs. However, considering this mechanism for WBAN reveals several problems for low-power in-body/on-body nodes such as pacemakers and defibrillator nodes/implants, which are not required to wake up periodically in order to exchange their schedules with other nodes. Furthermore, scheduled-contention mechanism may perform well for on-body applications but it does not provide reliable solutions to handle sporadic events including emergency and on-demand events. Handling sporadic events (emergency) requires innovative solutions that allow in-body/on-body nodes to update the coordinator within strictly limited amount of time.

\subsection{TDMA Mechanism}
In a TDMA mechanism, the channel is bounded by a superframe structure that consists of a number of time slots allocated by a base-station or a coordinator. The time slots are allocated according to the traffic requirements, i.e., a node gets a time slot whenever it has data to send or receive. Otherwise, it goes into sleep mode. Although it performs well in terms of power consumption but consumes extra energy due to frequent synchronization.

\begin{figure}[!h]
\centering
\includegraphics[width=3.5in]{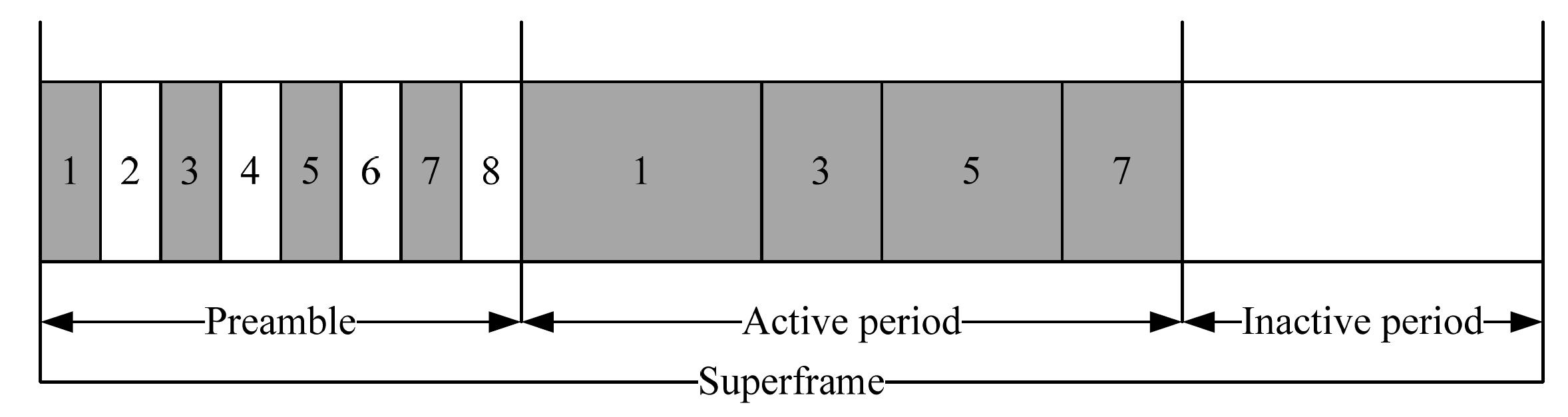}
\caption{PB-TDMA superframe structure}
\label{fig:7}
\end{figure}

The PB-TDMA protocol is based on the TDMA mechanism. In this protocol, the nodes are assigned specified slots for collision-free data transmission. These slots are repeated in fixed cycle. A complete cycle of these slots is called a frame. In the PB-TDMA protocol, each TDMA frame contains a preamble and a data transmission slot as illustrated in Fig. \ref{fig:7}. A node always listens to the channel during preamble and transmits in a data transmission slot. The preamble contains a dedicated subslot for every node. These subslots are used to activate the destination node by broadcasting its ID before transmission. After receiving the preamble, the destination node identifies the source node. Each node turns off its radio when it has no data to transmit. This mechanism avoids unnecessary power consumption of sensor nodes. The radio is turned on when the node finds its ID in the preamble or when the node has data to transmit.

As discussed in Section 2, the CSMA/CA protocol is not a reliable protocol for WBAN due to unreliable CCA, traffic correlation, and heavy collision problems. The alternative is to adapt a TDMA protocol that can solve the aforementioned problems in a power-efficient manner. However, traditional TDMA protocols such as PB-TDMA have several problems, i.e., preamble overhearing and limitation of handling sporadic events. Solving these problems (including many others) towards WBAN can accommodate the heterogeneous WBAN traffic in a power-efficient manner. Furthermore, new techniques are required to accommodate sporadic events in a reliable way.

\subsection{Comparison of LPL, Scheduled-contention and TDMA Mechanisms for WBAN}

Table \ref{tab:6} presents the characteristics of the LPL, schedule-contention, and TDMA mechanisms for WBAN [20]. The table shows that LPL and scheduled-contention are unable to accommodate the heterogeneous WBAN traffic including sporadic events. Although it is possible to develop new MAC protocols based on these mechanisms, they will not be able to satisfy all the requirements. For example, LPL mechanisms may perform well in case of periodic traffic but they are unable to accommodate aperiodic (unpredictable sporadic events) traffic and low duty cycle nodes. Furthermore, the scheduled-contention mechanisms are unable to cover in-body nodes, which do not require frequent synchronization or exchange of their schedules. The TDMA mechanisms provide good solutions to the variable WBAN traffic. The slots can be assigned according to the traffic volume of a node. Although in traditional TDMA protocols, nodes are required to synchronize at the beginning of each superframe boundary, but this approach can be optimized for nodes that do not require frequent synchronization. One way is to skip the synchronization control packets such as the beacon. The control packets (beacons) can be received when the low duty cycle nodes have data to send or receive. A detailed comparison of MAC protocols based on LPL, scheduled-contention, and TDMA mechanisms for WBAN is given in Table \ref{tab:7}\footnote{The protocols given in the table are not explained here due to space limitation problems. The table is a result of a comprehensive study of these protocols in the context of WBAN}.

\begin{table*}[!t]
\renewcommand{\arraystretch}{1.3}
\caption{Comparison of LPL, Scheduled-contention, and TDMA mechanisms for WBAN}
\label{tab:6}
\centering
\begin{tabular}{|p{5cm}|p{5cm}|p{5cm}|}
\hline
LPL	& Scheduled-contention &	TDMA\\
\hline
10 times less expensive than listening for full contention period &	Listening for full contention period &	Low duty cycle \\
\hline
Asynchronous &	Synchronous	& Synchronous-Fine grained time synchronisation\\
\hline
Sensitive to tuning for neighbourhood size and traffic rate &	Sensitive to clock drift &	Very sensitive to clock drift \\
\hline
Poor performance when traffic rates vary greatly (Optimised for known periodic traffic) &	Improved performance with increase in traffic &	Limited throughput and number of active nodes \\
\hline
Receiver and polling efficiency is gained at the much greater cost of senders &	similar cost incurred by sender and receiver &	Require clustering\\
\hline
challenging to adapt LPL directly to new radios like IEEE 802.15.4 &	Scalable, adaptive, and flexible &	Limited scalability and adaptability to changes on number of nodes \\
\hline
Unable to accommodate aperiodic traffic (unpredictable sporadic events) and low duty cycle nodes in WBAN. Very hard to satisfy the WBAN traffic heterogeneity requirements &	Low duty cycle nodes do not require frequent synchronization/exchange of schedules in WBAN. Hard to satisfy the WBAN traffic heterogeneity requirements &	Low duty cycle nodes do not require frequent synchronization at the beginning of each superframe. Easy to satisfy the WBAN traffic heterogeneity requirements\\
\hline
\end{tabular}\end{table*}


\begin{table*}[!t]
\renewcommand{\arraystretch}{1.3}
\caption{Summary of existing MAC Protocols for WBAN}
\label{tab:7}
\centering
\begin{tabular}{|p{1.5cm}|p{1.2cm}|p{1.2cm}|p{4cm}|p{4cm}|p{4cm}|}
\cline{1-6}
Power-efficient Mechanisms & Protocols & Channels & Organization and Basic Operation & Advantages and Disadvantages & Adaptability to WBAN \\ \cline{1-6}

\multicolumn{1}{|p{1.5cm}|}{Low power Listening} & WiseMAC & 1 & Organized randomly and operation is based on listening & Scalable and adaptive to traffic load, Support mobility, low and high power consumption in low and high traffic conditions, and low delay & Good for high traffic applications, not suitable for low duty cycle in-body/on-body nodes    \\ \cline{2-6}
\multicolumn{1}{|p{1.5cm}|}{} & BMAC &	1	& Organized in slots and operation is based on schedules &	Flexible, high throughput, tolerable latency, and low-power consumption &	Good for high traffic applications    \\ \cline{2-6}
\multicolumn{1}{|p{1.5cm}|}{} & STEM &	2 &	Organized randomly having two sub-channels (control + data channel) and operation is based on wakeup schedules &	Suitable for events based applications &	Good for periodic traffic especially for low traffic applications. Suitable to handle sporadic events due to a separate control sub-channel. But hard to handle sporadic events when the traffic load is high \\ \cline{1-6}

\multicolumn{1}{|p{1.5cm}|}{Scheduled-contention} & SMAC &	1	& Organized in slots and operation is based on schedules &	High transmission latency, loosely synchronized, low throughput	& Good for high traffic applications. Suitable for applications where throughput is not a primary concern such as in-body medical applications  \\ \cline{2-6}
\multicolumn{1}{|p{1.5cm}|}{} & TMAC &	1	& Organized in slots and operation is based on schedules &	Queued packets are sent in a burst thus achieve better delay performance, loosely synchronized &	Good for high traffic applications. Early sleep problems allow the nodes to loose synchronization \\ \cline{2-6}
\multicolumn{1}{|p{1.5cm}|}{} & PMAC &	1	& Organized in hybrid mode and operation is based on listening &	Adaptation to changes might be slow, loosely synchronized, high throughput under heavy traffic & 	Good for delay-sensitive applications \\ \cline{2-6}
\multicolumn{1}{|p{1.5cm}|}{} & DMAC &	1 &	Organized in slots and operation is based on schedules &	better delay performance due to Sleep schedules, loosely synchronized, optimized for data forwarding sink	& On-body nodes can be prioritized according to their application requirements and a data tree can be built, where the WBAN coordinator can be a cluster node \\ \cline{1-6}

\multicolumn{1}{|p{1.5cm}|}{TDMA} & FLAMA &	1	& Organized in frames and operation is based on schedules	& Better end-to-end reliability and energy saving, smaller delays, high reliability &	Good for low-power applications. Adaptable to high traffic applications \\ \cline{2-6}
\multicolumn{1}{|p{1.5cm}|}{} & LEACH &	1	& Organized in clusters and operations is based on TDMA scheme	& Distributed protocol, no global knowledge required, extra overhead for dynamic clustering &	TDMA schedules should be created by the WBAN coordinator. Cluster head should not change (depending on minimum communication energy) as in the traditional LEACH  \\ \cline{2-6}
\multicolumn{1}{|p{1.5cm}|}{} & HEED & 1	& Organized in clusters and operations is based on TDMA scheme &	Good for energy efficiency, scalability, prolonged network lifetime, load balancing &	The WBAN coordinator acts as a cluster head. Unlike traditional HEED, the WBAN network size is often defined (by the physician)  \\ \cline{1-6}
 
\end{tabular}
\end{table*}

From the following table, it can be seen that LPL-based protocols such as WiseMAC and BMAC \cite{21} protocols are good for high traffic applications while STEM \cite{22} performs well for low traffic applications. Furthermore, STEM can also accommodate the sporadic events by using a separate control channel. However, increase in the traffic load decreases the probability of handling sporadic events. Schedule-contention protocols such as SMAC and TMAC \cite{23} are suitable for high traffic applications, PMAC \cite{24} for delay sensitive applications, and DMAC \cite{25} for priority-based applications (where the nodes have different priorities). As mentioned earlier, TDMA-based protocols can easily accommodate the heterogeneous WBAN traffic since they are adaptable to the traffic load, i.e., slots can be assigned according to the traffic volume. However, traditional TDMA-based protocols such as FLAMA \cite{26}, LEACH \cite{27}, and HEED \cite{28} are unable to satisfy the WBAN requirements as mentioned in the following table. In addition, Most of the existing MAC protocols are designed for a single channel only, i.e., they do not operate on Multi-PHYs simultaneously. The MAC transparency has been a hot topic for the MAC designers since different bands have different characteristics in terms of data rate, number of sub-channels in a particular frequency band/channel, and data prioritization. A good MAC protocol for WBAN should enable reliable operation on MICS, ISM, and UWB etc bands simultaneously. With regards to MICS band, the FCC has imposed several restrictions \cite{29}. According to the FCC

\emph{Within 5 seconds prior to initiating a communications session, circuitry associated with a medical implant programmer/control transmitter must monitor the channel (sub-channel) or channels (sub-channels) the MICS system devices intend to occupy for a minimum of 10 milliseconds per channel.}

\begin{figure}[!h]
\centering
\includegraphics[width=3.5in]{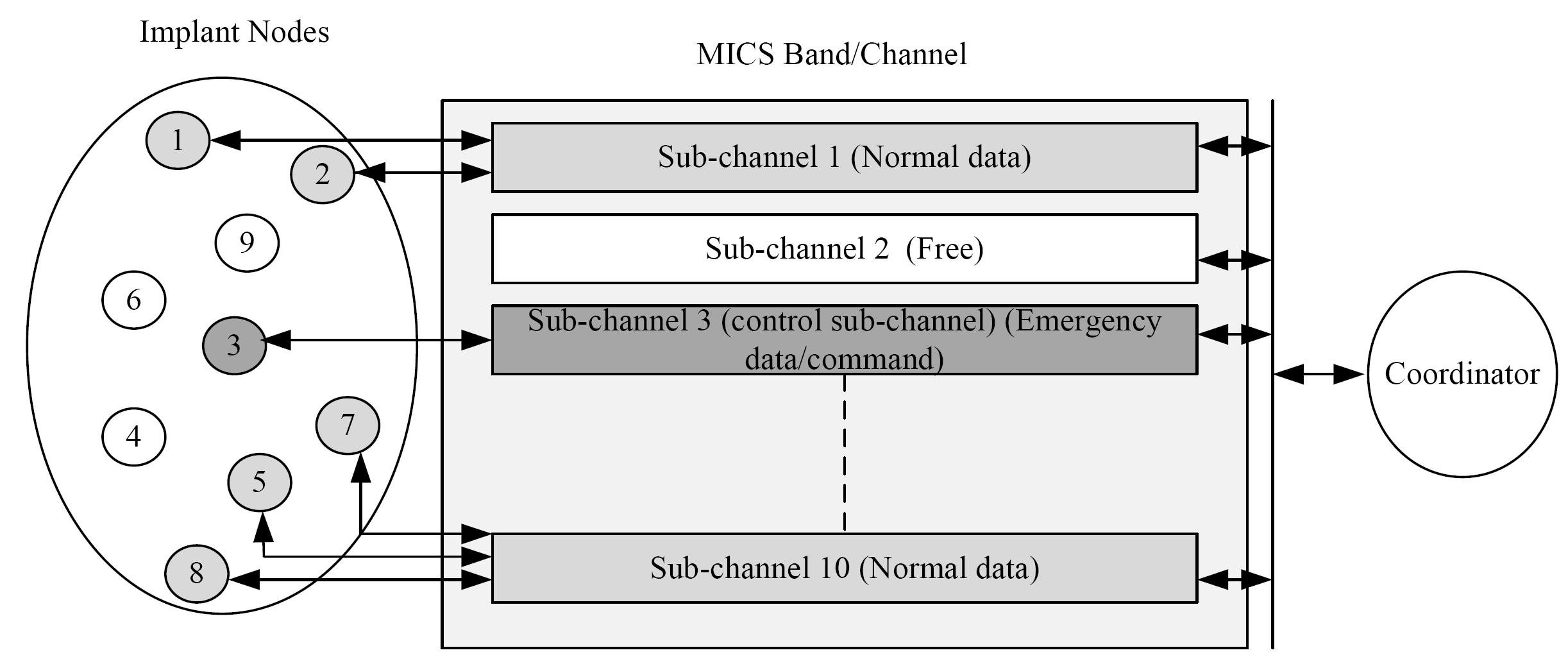}
\caption{An example of a control sub-channel for emergency traffic}
\label{fig:8}
\end{figure}

In other words, the coordinator must perform Listen-before-talking (LBT) criteria prior to establish a MICS communication sessions The sub-channels are solely assigned by the coordinator, i.e., the implants cannot initiate a communication session except in case of an emergency event. Furthermore, the implants are not allowed to perform LBT which creates problems to handle emergency events. Sending an emergency data regardless of LBT may result in heavy collision since the selected sub-channel may have data from another implant. The LBT restriction prevents MAC designers to develop a reliable mechanism for emergency traffic. One of the solutions is to use a wakeup radio or a control sub-channel dedicated temporarily to emergency traffic since the FCC does not allow the permanent dedication of a sub-channel in MICS band. The sub-channel information can be updated using control frames such as the beacon frame. Fig. \ref{fig:8} illustrates an example of a control sub-channel used to send emergency data. It shows that five nodes (1, 2, 5, 7 and 8) are transmitting normal data and one node (3) is transmitting emergency data. The remaining nodes (4, 6 and 9) are in sleep mode. Normal transmission requires beacon to allocate resources. While in emergency case, nodes are not required to wait for the beacon. The communication is initiated by the implants and a control sub-channel can be used to send the emergency data/command as given in the figure.

\section{conclusions}
This paper presented a comprehensive study of existing/proposed MAC protocols for WBAN with useful suggestions. Different power-efficient mechanisms such as LPL, schedule-contention, and TDMA mechanisms were analyzed and discussed in the context of WBAN. As discussed, the CSMA/CA protocol encounters heavy collision and unreliable CCA problems, hence the TDMA protocol was considered the most reliable and power-efficient protocol for WBAN. However, existing TDMA protocols have a number of limitations in terms of synchronization overhead, dynamic slot assignment, and Multi-PHYs communication.  Therefore, a novel low-power MAC protocol (probably based on a TDMA mechanism) is required to satisfy the traffic heterogeneity and correlation, MAC transparency, and reliability requirements. This study can be used as a guideline (for the beginners) towards the design and development of a new low-power MAC protocol for WBAN.

\section*{acknowledgement}
This research was supported by the The Ministry of Knowledge Economy (MKE), Korea, under the Information Technology Research Center (ITRC) support program supervised by the Institute for Information Technology Advancement (IITA) (IITA-2009-C1090-0902-0019).

%





\end{document}